\documentclass[a4paper,11pt]{article}

\usepackage[dvipdfmx]{graphicx}
\usepackage{amsthm} 
\usepackage[fleqn]{amsmath}
\usepackage[ruled]{algorithm2e}
\usepackage{amssymb,%
amsthm,%
color,%
colortbl,%
enumerate,%
MnSymbol,%
multirow,%
stmaryrd,%
url}

\newcommand{\PROBLEMNAMESTYLE}[1]{\textsc{#1}}
\newcommand{\COMPLEXITYCLASSNAMESTYLE}[1]{\mathsf{#1}}

\newcommand{\PROCEDURENAMESTYLE}[1]{\textsc{#1}}

\newcommand{\CEIL}[1]{\left\lceil{#1}\right\rceil}

\newcommand{\FLOOR}[1]{\left\lfloor{#1}\right\rfloor}

\newcommand{\ISSAT}{\textsc{IsSat}}

\newcommand{\MOD}{\,\mathrm{mod}\,}

\newcommand{\NP}{\COMPLEXITYCLASSNAMESTYLE{NP}}

\newcommand{\NUMERICALSEMIGROUP}[1]{NS}

\newcommand{\OT}{\PROBLEMNAMESTYLE{1In3}}

\newcommand{\PROOFSTARTX}[1]{{\noindent \itshape Proof #1.} }

\newcommand{\POS}{\PROBLEMNAMESTYLE{Pos}}

\newcommand{\QED}{\hfill$\square$}

\newcommand{\SAT}{\PROBLEMNAMESTYLE{Sat}}

\newcommand{\TDBS}{\PROCEDURENAMESTYLE{2DIBSearch}}

\newcommand{\ANBR}[1]{\langle{#1}\rangle}

\newcommand{\IVBR}[1]{\llbracket{#1}\rrbracket}

\newtheorem{definition}{Definition}
\newtheorem{claim}[definition]{Claim}

\newtheorem{lemma}[definition]{Lemma}
\newtheorem{observation}[definition]{Observation}
\newtheorem{problem}[definition]{Problem}

\DontPrintSemicolon%
\LinesNumbered%
\SetKwFor{For}{for}{}{}%
\SetKwFor{ForEach}{foreach}{}{}%
\SetKwFor{While}{while}{}{}%
\SetKwIF{If}{ElseIf}{Else}{if}{}{elseif}{else}{}%
\SetKwProg{Fn}{}{}{}

\title{Two-Dimensional Indirect Binary Search\\ 
for the Positive One-In-Three Satisfiability Problem}

\author{Shunichi Matsubara\thanks{matsubara@it.aoyama.ac.jp}\\
Aoyama Gakuin University,\\
5-10-1, Fuchinobe, Chuo-ku, Sagamihara, Kanagawa, \\
252-5258, Japan
}
\date{}

\begin{document}

\maketitle

\begin{abstract}
In this paper,
we propose an algorithm 
for the positive-one-in-three satisfiability problem ($\POS\OT\SAT$).
The proposed algorithm decides the existence of a satisfying assignment 
in all assignments for a given formula
by using a $2$-dimensional binary search method
without constructing an exponential number of assignments.
\end{abstract}

\section{Introduction}
\label{sec:introduction}

In this paper,
we propose an
algorithm for the positive-one-in-three satisfiability problem ($\POS\OT\SAT$).
$\POS\OT\SAT$ is known to be $\NP$-complete~\cite{Schaefer:1978:CSP:800133.804350}.
We prove that the proposed algorithm can run efficiently.

The proposed algorithm decides whether there is a satisfying assignment 
for a given positive $3$CNF formula
by using a $2$-dimensional version of the binary search method.
First, it constructs an equivalent positive $3$CNF formula 
for the given formula as the preprocess for the binary search.
Then, 
it encodes all partial assignments to single variables in the constructed positive $3$CNF formula.
As a result,
we obtain a matrix whose components means a truth assignment in the constructed formula.
The algorithm does the binary search for that matrix.
Every row and column in that matrix are sorted in ascending order.
Thus, the algorithm can expectedly do the binary search.
Representing all components of the matrix 
requires an exponential space for the size of the input formula.
However, we can use the matrix without constructing all the components.

In Section~\ref{sec:preliminary},
we define some basic concepts and notation.
In Section~\ref{sec:algorithm},
we propose an algorithm for $\POS\OT\SAT$.
Then, we prove its validity and analyze its running time.

\section{Basic concepts and notation}
\label{sec:preliminary}

In this section, 
we define basic concepts and notation
that are used throughout the paper.
We follow convention of literature in theoretical computer science or combinatorics.

We denote the empty string by $\varepsilon$.
We denote by $\llbracket p \rrbracket$ 
a characteristic function on a predicate $p$; i.e.,
$\IVBR{p}$ is $1$ if $p = 1$,
and $0$ otherwise.

We denote the sets of all nonnegative and positive integers 
by $\mathbb{N}$ and $\mathbb{N}_+$, respectively.
Given $l,u \in \mathbb{N}$,
we denote the interval
$\{i \in \mathbb{N} \colon l \leq i \leq u\}$
by $[l,u]$.
Given $l, u \in \mathbb{N}$ 
and $f \colon \mathbb{N} \rightarrow \mathbb{N}$,
if $l > u$, 
then we consider $\sum_{x=l}^u f(x)$ to be $0$.

Let $\Phi$ be a finite set.
Let $n \in \mathbb{N}_+$.
We denote a vector in $\Phi^n$ by a lower case bold symbol.
We denote $\bigcup_{i \in \mathbb{N}_+} \Phi^i$ by $\Phi^\ast$. 
Let $\mathbf{b} \in \Phi^n$.
Given $\mathbf{b}$, for every $i \in [1, n]$,
we represent the $i$th component of $\mathbf{b}$
by the corresponding normal weight symbol $b_i$; 
i.e., $\mathbf{b} = (b_1, \cdots, b_n)$.
Conversely, given $n$ elements $b_1, \cdots, b_n$ of $\Phi$,
$\mathbf{b}$ denotes the vector $(b_1, \cdots, b_n)$.
Given $l, u \in \mathbb{N}$,
we denote the vector $(b_l, \cdots, b_u)$ by $\mathbf{b}_{l:u}$.
Given $\delta \in \Phi$ and $k \in \mathbb{N}_+$,
we denote the vector $\underbrace{(\delta, \cdots, \delta)}_{k}$
by ${\boldsymbol \delta}_{(k)}$.
We omit the subscript {\itshape ``$(k)$''} if no confusion arises.
Let $\mathbf{a} \in \Phi^n$.
We denote $\mathbf{a} \leq \mathbf{b}$ 
if and only if
$a_i \leq b_i$ for every $i \in [1, n]$.
Given $\mathbf{a}, \mathbf{b}$,
we denote the inner product $\sum_{i=1}^n a_i b_i$
by $\mathbf{a} \cdot \mathbf{b}$.
For convenience,
we identify a vector in $\Phi^n$
with a sequence of length $n$ over $\Phi$,
a string of length $n$ over $\Phi$,
or a mapping from $[1, n]$ to $\Phi$
if no confusion arises.
For example,
we identify the vector $(b_1, \cdots, b_n)$
with the sequence $b_1, \cdots, b_n$,
the string $b_1 \cdots b_n$,
or a mapping that maps $i$ to $b_i$ for every $i \in [1, n]$.
We denote $b \in \mathbf{b}$ if $\mathbf{b}$ contains $b$ as a component.
We define a binary relation $\subseteq$ over $\Phi^\ast$
as follows. 
Given $\mathbf{a}$ and $\mathbf{b}$ in $\Phi^n$,
$\mathbf{a} \subseteq \mathbf{b}$
if and only if $a_i = b_i$ for every $i \in [1, |\mathbf{a}|]$.
Given $\mathbf{b}$,
we denote its reverse $(b_n, \cdots, b_1)$
by $\mathbf{b}^{\mathrm{R}}$.
Given $l, u \in [1, n]$,
$\mathbf{b}_{l:u}^{\mathrm{R}}$
denotes $(\mathbf{b}_{l:u})^{\mathrm{R}}$.

We say that a finite set $A$ of integers can be $2$-dimensionally-sorted 
in ascending (descending) order
if there is a matrix $M$ such that 
every row and column are sorted in ascending (descending) order;
and there is some one to one correspondence
from $A$ to the set of all components of $M$.

\subsection{Concepts and notation on integers}
\label{subsec:preliminary_integer}

Let $b \in \mathbb{N}_+$. 
Let $n, k \in \mathbb{N}$ 
such that $\FLOOR{\log_b n} + 1 \leq k$.
If $d_i$ is $\left(\FLOOR{n/b^{i-1}} \MOD b\right)$
for every $i \in [1, k]$,
then we call the string $d_k \cdots d_1$
over $[0, b-1]$ the {\itshape base-$b$ representation} of $n$ of length $k$.
We omit the phrase {\itshape ``of length $k$''}
if no confusion arises.
Given a base-$b$ representation $\alpha$ of some $m \in \mathbb{N}$,
$(\alpha)_b$ denotes the integer $m$. 
We consider $\varepsilon$ to be the base-$b$ representation of $0$ of length $0$;
i.e., we consider $(\varepsilon)_b$ to be $0$.
Given $l, u \in [1, k]$ with $l < u$,
if $d_k \cdots d_1$ is the base-$b$ representation of $n$,
then we call the substring $d_l \cdots d_u$
the {\itshape base-$b$ $(l, u)$-zone of $n$}.
We omit the phrase {\itshape ``base-$b$''}
if no confusion arises.

\subsection{Boolean formulae}
\label{subsec:preliminary_boolean_formulae}

In this subsection,
we define notation and assumptions and review some concepts 
on Boolean formulae.
We assume the reader to be familiar to basic concepts in Boolean satisfiability.
The reader is referred to 
some books by 
some chapters in 
Arora and Barak~\cite{arora2009computational}, 
Creignou, Khanna, and Sudan~\cite{creignou2001complexity},
or Wegener~\cite{Wegener:1987:CBF:35517}
if necessary.

\subsubsection{Assumptions}
\label{subsubsec: Boolean_formula_assumptions}

In this subsubsection, 
we define assumptions on Boolean formulae,
which we use throughout the paper.
These assumptions are for technical reasons.
By those assumption,
we do not lose the generality of discussion
on polynomial-time computability.

We fix $Z$ to be a countable set of Boolean variables.
For every $i \in \mathbb{N}_+$
$z_i$ denotes a Boolean variable in $Z$.
We assume that every Boolean formula in this paper
is defined over $Z$.
We fix $\varphi$ and $\psi$ to be positive $3$CNF formulae over $Z$.

We assume a clause in a Boolean formula to be a sequence of literals
although a clause is often assumed to be a set of literals in other literature.
For example, we distinguish $z_1 \lor z_2 \lor z_3$ 
from $z_3 \lor z_2 \lor z_2$.
Similarly,
we assume a CNF formula to be a sequence of clauses
although a CNF formula is often assumed to be a set of clauses in other literature.
For example, given clauses $C_1$ and $C_2$,
we distinguish a conjunction $C_1 \land C_2$ from $C_2 \land C_1$.
Needless to say,
the satisfiability of a given formula
do not depend on
whether clauses or formulae are regarded as sets or sequences.

Let $\varphi$ be a given.
We assume that $\varphi$ consists of $2$ or more clauses.
We assume that every clause in $\varphi$ contains distinct variables.
We assume that no two clauses in $\varphi$ 
consist of the identical combination of variables.
We assume that 
the indices of variables occurring in $\varphi$
are successive integers from $1$;
i.e., the set of all variables occurring in $\varphi$
can be represented as $\{z_i \colon i \in [1, k]\}$ for some $k \in \mathbb{N}$.
We consider the size of a Boolean formula to be the number of variables in the formula.

\subsubsection{Concepts and notation}

Let $z \in Z$.
Then, we call $z$ or $\lnot z$ literals.
In particular,
we call $z$ a positive literal;
and $\lnot z$ a negative literal.
We say that $\varphi$ is {\itshape positive} 
if $\varphi$ consists of only positive literals.
We denote the set of all variables in $\varphi$ by $V(\varphi)$.
We denote the set of all clauses in $\varphi$ by $\mathcal{C}_\varphi$.

Suppose that $\varphi$ is represented
as $C_m \land \cdots \land C_1$;
i.e., $m = |\mathcal{C}_\varphi|$.
Let $k$ be the number of variables in $\varphi$.
We define a {\itshape partial assignment} ${\boldsymbol \sigma}$ for $\varphi$
as a mapping from $\{z_i \colon i \in [1, j]\}$ to $\{0, 1\}$, 
where $j \in [1, |V_\varphi|]$.  
We call ${\boldsymbol \sigma}$ a {\itshape truth assignment} for $\varphi$
if $|{\boldsymbol \sigma}| = |V(\varphi)|$.
We often call a truth assignment for $\varphi$ simply an assignment for $\varphi$.
Let ${\boldsymbol \sigma} \in \{0, 1\}^\nu$.
We say that a partial assignment ${\boldsymbol \sigma}$ is 
a {\itshape $1$-in-$3$-satisfying} for $\varphi$
if $|\{z \colon {\boldsymbol \sigma}(z) = 1, z \in C_j \}| \leq 1$
for every $j \in [1, m]$.
We say that an assignment ${\boldsymbol \sigma}$ 
is {\itshape $1$-in-$3$-satisfying} for $\varphi$
if $|\{z_i \colon {\boldsymbol \sigma}(z_i) = 1 \}| = 1$
for every $j \in [1, m]$.
Given $\varphi$,
if there is a $1$-in-$3$-satisfying assignment ${\boldsymbol \sigma}$,
then we say that $\varphi$ is $1$-in-$3$-satisfiable.
We consider $\varepsilon$ to be a partial assignment that does not assign anything.
Given a partial assignment ${\boldsymbol \sigma}$ for $\varphi$ 
and a literal $z$ in $\varphi$,
we call $z$ a {\itshape true literal} in ${\boldsymbol \sigma}$
if $\sigma(z) = 1$.
Let $b \in \mathbb{N}_+$.
Given $i \in [1, k]$,
we denote the integer $\sum_{j=1}^m b^{i-1} \IVBR{z_i \in C_j}$
by $(\varphi, z_i)_b$.
We denote the vector 
$((\varphi, z_1)_4, \cdots, (\varphi, z_k)_4)$
by $\widehat{{\boldsymbol \varphi}}$.

\begin{problem}[$\POS\OT\SAT$] \
\label{prob:1}

\noindent{\itshape Instance.} 
A positive $3$CNF formula $\varphi$.

\noindent{\itshape Question.} 
Is $\varphi$ $1$-in-$3$-satisfiable?

\end{problem}

\subsection{Computational complexity}
\label{subsec:computational_complexity}

We assume the reader to be familiar to basic concepts and results 
in computational complexity theory.
The reader is referred to Arora and Barak~\cite{arora2009computational}
if necessary.
Basically,
we estimate the running time of an algorithm
by using a function in the bit length of a given input.
On the other hand,
this paper focuses on the polynomial-time computability 
of $\POS\OT\SAT$.
Thus, we analyze the running time of an algorithm
roughly to some extent that we do not lose the correctness
in favor of clarity of discussion.
For example, 
as we described in Subsubsection~\ref{subsubsec: Boolean_formula_assumptions},
we adopt the number of variables 
as the size of a given $3$CNF formula.

\section{Algorithm}
\label{sec:algorithm}

In this section,
we propose a new algorithm for $\POS\OT\SAT$.
Subsection~\ref{subsec:ideas_of_algorithm} 
describes the outline and key ideas of this algorithm informally.
Subsection~\ref{subsec:details_of_algorithm},
describes the details of the algorithm formally.
In Subsection~\ref{subsec:validity_of_algorithm},
we prove the validity of the algorithm.
Finally, 
in Subsection~\ref{subsec:running_time_of_algorithm},
we analyze the running time of the algorithm.

For preparation, 
we fix some symbols as follows. 
$C_{m_1} \land \cdots \land C_1$
denotes a positive $3$CNF formula $\psi$.
Moreover,
$k_1$ denotes the number of variables in $\psi$.
For every $i \in [1, m_1]$ and $j \in [1, 3]$,
$\ANBR{i, j; \psi}$ denotes an integer in $[1, k_1]$
such that $C_i = z_{\ANBR{i, 3; \psi}} \lor z_{\ANBR{i, 2; \psi}} \lor z_{\ANBR{i, 1; \psi}}$.
We denote $\ANBR{i, j; \psi}$ by simply $\ANBR{i, j}$
if no confusion arises.

\subsection{Ideas}
\label{subsec:ideas_of_algorithm}

In this subsection,
we first outline the algorithm that we propose in this paper.
After that, we describe some intuitive ideas of the algorithm
by executing the algorithm for a $3$CNF formula.

\subsubsection{Outline}
\label{subsubsec:algorithm_outline}

In the proposed algorithm,
every Boolean variable $z_i$ in a given $\psi$
is encoded to $(\psi, z_i)_4$.
Given $(\psi, z_i)_4$,
we can observe the following meaning
for its base-$4$ representation of length $m_1$. 
In that base-$4$ representation,
the $j$th digit from the right end means 
whether the clause $C_j$ contains $z_i$,
where $j \in [1, m_1]$.
That is,
$(\psi, z_i)_4$ simulates the assignment of $1$ to $z_i$ in $\psi$.
Thus, we can represent a total truth assignment for $\psi$
as the integer $\sum_{i \in I} (\psi, z_i)_4$
for some $I \subseteq [1, k_1]$.
If $i \in I$, then we consider $z_i$ to be assigned $1$ in $\psi$.
Then, a satisfying assignment for $\psi$ corresponds to $(\underbrace{1 \cdots 1}_{m_1})_4$;
i.e., $\sum_{i=1}^{m_1} 4^{i-1}$.

A basic strategy in the algorithm 
is a $2$-dimensional version of binary search.
First, the algorithm does a preprocess for the given $\psi$.
By that procedure,
we construct a positive $3$CNF formula $\varphi$ of $m$ clauses and $k$ variables.
Then, the algorithm searches the integer $\sum_{i=1}^{m} 4^{i-1}$
in the set of integers
$\sum_{i \in I_0} (\psi, z_i)_4, \cdots, \sum_{i \in I_\alpha} (\psi, z_i)_4$,
where $\alpha = 2^{k} - 1$ and $I_0, \cdots, I_\alpha$ are distinct subsets of $[1, k]$.
Needless to say,
an exponential space is necessary
to explicitly construct all the integers
$\sum_{i \in I_0} (\psi, z_i)_4, \cdots, \sum_{i \in I_\alpha} (\psi, z_i)_4$.
Thus, 
we can do this search without explicitly constructing the overall sequence.
Moreover,
the sequence 
$\sum_{i \in I_0} (\psi, z_i)_4, \cdots, \sum_{i \in I_\alpha} (\psi, z_i)_4$
is required to be sorted in an order.
Sorting these integers in $1$-dimension
appears to be difficult.
However,
if we arrange those integers in $2$-dimension,
then we can sort them, as we will describe below.

We fix $\psi_1$ 
to be a positive $3$CNF
$\underbrace{(z_{1} \lor z_{2} \lor z_{3})}_{C_2} 
\land \underbrace{(z_{1} \lor z_{2} \lor z_{4})}_{C_1}$.
In the remaining part of this subsection,
we will describe the details of the proposed algorithm for $\psi_1$.
In the algorithm, for convenience,
we replace $\psi_1$ by new $3$CNF formulae some times.
Thus, for every symbol,
we often use a parenthesized superscript 
for distinguishing the phase when the symbol is used.

\subsubsection{Preprocess}
\label{subsubsec:example_of_preprocess}

As a preparation for the main search,
the proposed algorithm construct a new $3$CNF formula from $\psi_1$,
and then encodes it to a set of integers. 
Let us describe it in more detail below.
Let us represent $\psi_1$ in the earliest phase of the algorithm
by 
$\underbrace{(z_{1}^{(0)} \lor z_{2}^{(0)} \lor z_{3}^{(0)})}_{C_{2}^{(0)}} 
\land \underbrace{(z_{1}^{(0)} \lor z_{2}^{(0)} \lor z_{4}^{(0)})}_{C_{1}^{(0)}}$.
Then, for every $i \in [1, 4]$,
the base-$4$ representation for $(\psi_1^{(0)}, z_i^{(0)})_4$ 
is illustrated in Table~\ref{tab:2}.
\begin{table}[htbp]
\begin{center}
\caption{Base-$4$ representation for $(\psi_1^{(0)}, z_i^{(0)})_4$, where $i \in [1, 4]$.}
\label{tab:2}
\begin{tabular}{ccccc}
\hline
& $(\psi_1, z_1)_4$
& $(\psi_1, z_2)_4$
& $(\psi_1, z_3)_4$
& $(\psi_1, z_4)_4$
\\
\hline
Base-$4$ representation & $11$ & $11$ & $10$ & $01$
\\
\hline
 \end{tabular}
\end{center}
\end{table}

First, the algorithm replaces the indices of the variables
so that $(\psi_1, z_1)_4 \leq (\psi_1, z_2)_4  \leq (\psi_1, z_3)_4 \leq (\psi_1, z_4)_4$.
Let us represent $z_i$ as $z_i^{(1)}$ for every $i \in [1, 4]$
and $C_j$ as $C_j^{(1)}$ for every $j \in \{1, 2\}$
in the phase immediately after those replacements.
That is, in this phase,
$\psi^{(1)} = \underbrace{(z_{4}^{(1)} \lor z_{3}^{(1)} \lor z_{2}^{(1)})}_{C_{2}^{(1)}} 
\land \underbrace{(z_{4}^{(1)} \lor z_{3}^{(1)} \lor z_{1}^{(1)})}_{C_{1}^{(1)}}$.
In $\psi^{(1)}$, 
every occurrence of every variable  can be represented 
as Table~\ref{tab:1}.

\begin{table}[htbp]
\caption{Correspondence between the indices of variables in two ways.}
\label{tab:1}
\begin{center}
\begin{tabular}{ccccccc}
\hline
& $z_{\ANBR{2,3}}$ & $z_{\ANBR{2,2}}$ & $z_{\ANBR{2,1}}$ 
& $z_{\ANBR{1,3}}$ & $z_{\ANBR{1,2}}$ & $z_{\ANBR{1,1}}$ 
\\
\hline
& $z_4$ & $z_3$ & $z_2$ & $z_4$ & $z_3$ & $z_1$  
\\
\hline 
\end{tabular}
\end{center}
\end{table}

In the next phase,
the algorithm constructs three clauses 
for every clause $C_j^{(1)}$, where $j \in \{1,2\}$,
by using the variables $z_{\ANBR{j, 1}}$, $z_{\ANBR{j, 2}}$, and $z_{\ANBR{j, 3}}$
and new variables 
$z_{k_1 + 4(j-1) + 1}$, $z_{k_1 + 4(j-1) + 2}$, and $z_{k_1 + 4(j-1) + 3}$.
In this phase,
let us use ``$(2)$'' as a superscript of every symbol.
In more details, we construct the following clauses.
\begin{eqnarray*}
&&
C_{4j}^{(2)} = (z_{k_1 + 4(j-1) + 3} \lor z_{k_1 + 4(j-1) + 2} \lor z_{\ANBR{j, 3}})\\ 
&&
C_{4j-1}^{(2)} = (z_{k_1 + 4(j-1) + 3} \lor z_{k_1 + 4(j-1) + 1} \lor z_{\ANBR{j, 2}}) \\ 
&&
C_{4j-2}^{(2)} = (z_{k_1 + 4(j-1) + 3} \lor z_{\ANBR{j, 3}} \lor z_{\ANBR{j, 2}}) 
\end{eqnarray*}
Moreover, the algorithm renames the clause $C_j$ as $C_{4j-3}$;,
i.e., $C_{4j-3}^{(2)} = C_j^{(1)}$.
That is, the algorithm constructs the following $\psi_1^{(2)}$.

\begin{eqnarray*}
&& 
\underbrace{(z_{10} \lor z_{9} \lor z_{4})}_{C_8} 
\land \underbrace{(z_{10} \lor z_{8} \lor z_{3})}_{C_7}
\land \underbrace{(z_{10} \lor z_{4} \lor z_{3})}_{C_6}
\land \underbrace{(z_{4} \lor z_{3} \lor z_{2})}_{C_5} 
\\
&&
\land \underbrace{(z_{7} \lor z_{6} \lor z_{4})}_{C_4}  
\land \underbrace{(z_{7} \lor z_{5} \lor z_{3})}_{C_3}  
\land \underbrace{(z_{7} \lor z_{4} \lor z_{3})}_{C_2} 
\land \underbrace{(z_{4} \lor z_{3} \lor z_{1})}_{C_1} 
\end{eqnarray*}

\subsubsection{Sorted matrix}
\label{subsubsec:TDS_sequence}

After the preprocess in Subsubsection~\ref{subsubsec:example_of_preprocess},
we can find a $(2^4 \times 2^6)$-matrix $M_{\psi_1}$, 
each of whose rows and columns is sorted.
In this subsubsection,
we describe more details of $M_{\psi_1}$.
In this subsubsection, 
we fix ${\boldsymbol \sigma}$ to be an assignment for $\psi_1^{(2)}$.
Every component in $M_{\psi_1}$ corresponds to an assignment for $\psi_1$;
and conversely,
given an assignment ${\boldsymbol \sigma}$, 
there is a component in $M_{\psi_1}$ corresponding to ${\boldsymbol \sigma}$.
Given ${\boldsymbol \sigma}$,
we denote $(({\boldsymbol \sigma}_{1:4}^R)_2+1, ({\boldsymbol \sigma}_{5:10}^R)_2+1)$
by $f_{\psi_1}({\boldsymbol \sigma})$.
The pair $f_{\psi_1}({\boldsymbol \sigma})$ means the position of a component in $M_{\psi_1}$.
Then,
$M_{\psi_1}$ is a matrix
such that every row and column are sorted in ascending order.

\begin{table}[htbp]
\caption{$(\psi_1^{(2)}, z_i^{(2)})_4$ for every $i \in [1, 4]$.}
\label{tab:3}
\begin{center}
\begin{tabular}{cccc}
\hline
$i$ 
& ${\boldsymbol \sigma^{\mathrm{R}}}$
& $f_{\psi_1}({\boldsymbol \sigma})$
& $(\psi_1, z_{i})_{4}$
\\
\hline
$1$ & $0000000001$ & $((0001)_2, 0)$ & $(00000001)_{4}$ \\
$2$ & $0000000010$ & $((0010)_2, 0)$ & $(00010001)_{4}$ \\
$3$ & $0000000100$ & $((0100)_2, 0)$ & $(01120112)_{4}$ \\
$4$ & $0000001000$ & $((1000)_2, 0)$ & $(11231123)_{4}$ \\
\hline
\end{tabular}
\end{center}
\end{table}

By Tables~\ref{tab:3}~and~\ref{tab:4},
we can observe that the larger $i$ is, the larger $(\psi, z_i)_4$ is.
Table~\ref{tab:3} and Table~\ref{tab:4}
show the constructed integers that affect the ordering of the magnitudes  
in the column and row directions in $M_{\psi_1}$, respectively.

\begin{table}[htbp]
\caption{Integer ${\boldsymbol \sigma} \cdot \widehat{{\boldsymbol \psi}}_1$,
which is the value of the $f_{\psi_1}({\boldsymbol \sigma})$-component in $M_{\psi_1}$,
for ${\boldsymbol \sigma} \in \{1\}^i \{0\}^{10-i}$,
where $i \in [1, 4]$.}
\label{tab:5}
\begin{center}
 \begin{tabular}{cccc}
 \hline
$i$ 
& ${\boldsymbol \sigma^{\mathrm{R}}}$
& $f_{\psi_1}({\boldsymbol \sigma})$
& ${\boldsymbol \sigma} \cdot \widehat{{\boldsymbol \psi}}_1$
 \\
 \hline
$1$ & $0000000001$ & $((0001)_2, 0)$ & $(00000001)_{4}$ \\
$2$ & $0000000011$ & $((0011)_2, 0)$ & $(00010001)_{4}$ \\
$3$ & $0000000111$ & $((0111)_2, 0)$ & $(01120112)_{4}$ \\
$4$ & $0000001111$ & $((1111)_2, 0)$ & $(11231123)_{4}$ \\ \hline
\end{tabular}
\end{center}
\end{table}

\begin{table}[htbp]
\caption{$(\psi_1^{(2)}, z_i^{(2)})_4$ for every $i \in [5, 10]$.}
\label{tab:4}
\begin{center}
\begin{tabular}{cccc}
\hline
$i$ 
& ${\boldsymbol \sigma^{\mathrm{R}}}$
& $f_{\psi_1}({\boldsymbol \sigma})$
& $(\psi_1, z_{i})_{4}$
\\
\hline
$5$ & $0000010000$ & $(0, (000001)_2)$ & $(00000100)_{4}$ \\
$6$ & $0000100000$ & $(0, (000010)_2)$ & $(00001100)_{4}$ \\
$7$ & $0001000000$ & $(0, (000100)_2)$ & $(00002210)_{4}$ \\
$8$ & $0010000000$ & $(0, (001000)_2)$ & $(01002210)_{4}$ \\
$9$ & $0100000000$ & $(0, (010000)_2)$ & $(11002210)_{4}$ \\
$10$ & $1000000000$ & $(0, (100000)_2)$ & $(22102210)_{4}$ \\
\hline
\end{tabular}
\end{center}
\end{table}
\begin{table}[htbp]
\caption{Integer ${\boldsymbol \sigma} \cdot \widehat{{\boldsymbol \psi}}_1$,
which is the value of the $f_{\psi_1}({\boldsymbol \sigma})$-component in $M_{\psi_1}$
for ${\boldsymbol \sigma} \in \{1\}^i \{0\}^{10-i}$,
where $i \in [5, 10]$.}
\label{tab:6}
\begin{center}
\begin{tabular}{cccc}
\hline
$i$ 
& ${\boldsymbol \sigma^{\mathrm{R}}}$
& $f_{\psi_1}({\boldsymbol \sigma})$
& ${\boldsymbol \sigma} \cdot \widehat{\boldsymbol \psi}_1$
\\
\hline
$5$ & $0000010000$ & $(0, (000001)_2)$ & $(00000100)_{4}$ \\
$6$ & $0000110000$ & $(0, (000011)_2)$ & $(00001100)_{4}$ \\
$7$ & $0001110000$ & $(0, (000111)_2)$ & $(00002210)_{4}$ \\
$8$ & $0011110000$ & $(0, (001111)_2)$ & $(01002210)_{4}$ \\
$9$ & $0111110000$ & $(0, (011111)_2)$ & $(11002210)_{4}$ \\
$10$ & $1111110000$ & $(0, (111111)_2)$ & $(22102210)_{4}$ \\
\hline
\end{tabular}
\end{center}
\end{table}

Table~\ref{tab:5} shows that
the first row in $M_{\psi_1}$ is sorted.
By that table, 
we can find that every row in $M_{\psi_1}$ is sorted. 
Table~\ref{tab:5} shows
that the first column in $M_{\psi_1}$ is sorted.
By that table, 
we can find that every column in $M_{\psi_1}$ is sorted.
By Tables~\ref{tab:5} and \ref{tab:6}, 
we obtain the following property.

\begin{observation}
\label{obs:2}
Let $(l_1, u_1) \in [1, 3] \times [5, 9]$.
Let $(l_2, u_2) \in [l_1+1, 4] \times [u_1+1, 10]$.
Then,
\begin{eqnarray*}
&&
2^4 \sum_{i=5}^{u_1} (\psi_1^{(2)}, z_i^{(2)})_4
+ \sum_{i=1}^{l_1} (\psi_1^{(2)}, z_i^{(2)})_4
\ \ \leq \ \
2^4 (\psi_1^{(2)}, z_{u_2}^{(2)})_4
+ (\psi_1^{(2)}, z_{l_2}^{(2)})_4.
\end{eqnarray*}
\end{observation}

By Observation~\ref{obs:2}, the following holds.
\begin{observation}
\label{obs:3}
Let ${\boldsymbol \sigma}$ and ${\boldsymbol \mu}$ be 
vectors such that
$f_{\psi_1}({\boldsymbol \sigma}) \leq f_{\psi_1}({\boldsymbol \mu})$.
Then,
\begin{eqnarray*}
{\boldsymbol \sigma} \cdot \widehat{\boldsymbol \psi}_1
\leq {\boldsymbol \mu} \cdot \widehat{\boldsymbol \psi}_1.
\end{eqnarray*}
\end{observation}

\begin{figure}[htbp]
\begin{center}
\includegraphics[width=15cm]{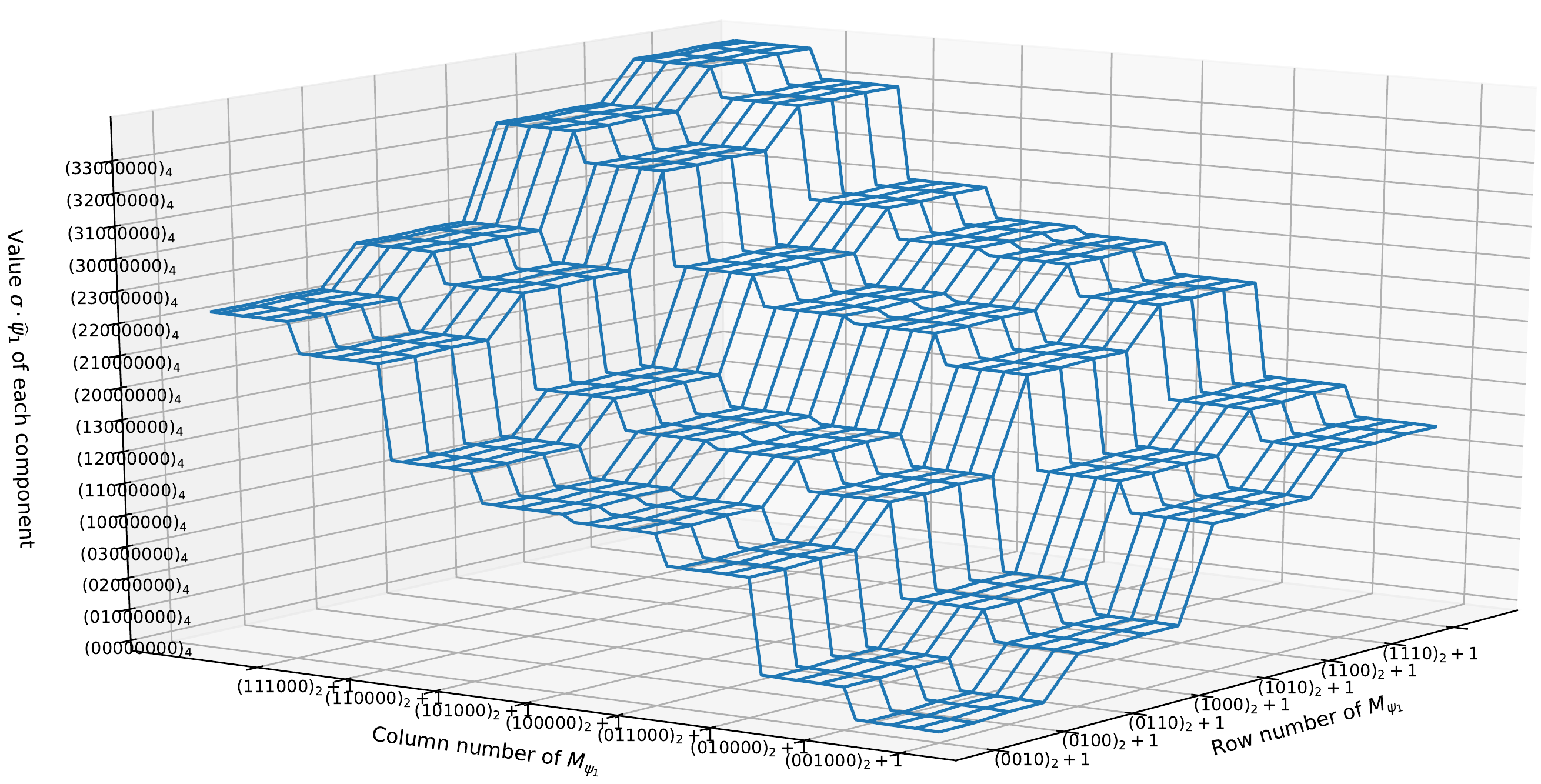}
\caption{Values of components of the matrix $M_{\psi_1}$,
whose $(i, j)$-element is $f^{-1}(i-1, j-1) \cdot \widehat{\boldsymbol \psi}_1$,
where $i \in [1, 2^4]$ and $j \in [1, 2^6]$.}
\label{fig:matrix_M_psi1_in_3d} 
\end{center}
\end{figure}

Figure~\ref{fig:matrix_M_psi1_in_3d} 
visualizes the matrix $M_{\psi_1}$.
We can see Observation~\ref{obs:3} in Figure~\ref{fig:matrix_M_psi1_in_3d}.

Consequently, we find the following observation.
\begin{observation}
\label{obs:4}
Let $M_{\psi_1}$ be the matrix
whose $(i, j)$-element is $f^{-1}(i-1, j-1) \cdot \widehat{\boldsymbol \psi}_1$,
where $i \in [1, 2^4]$ and $j \in [1, 2^6]$.
Then, $M_{\psi_1}$ is sorted in ascending order.
\end{observation}

\subsubsection{Indirect search for an implicit matrix}
\label{subsubsec:implicit_serarch}

\begin{figure}[htbp]
\begin{center}
\includegraphics{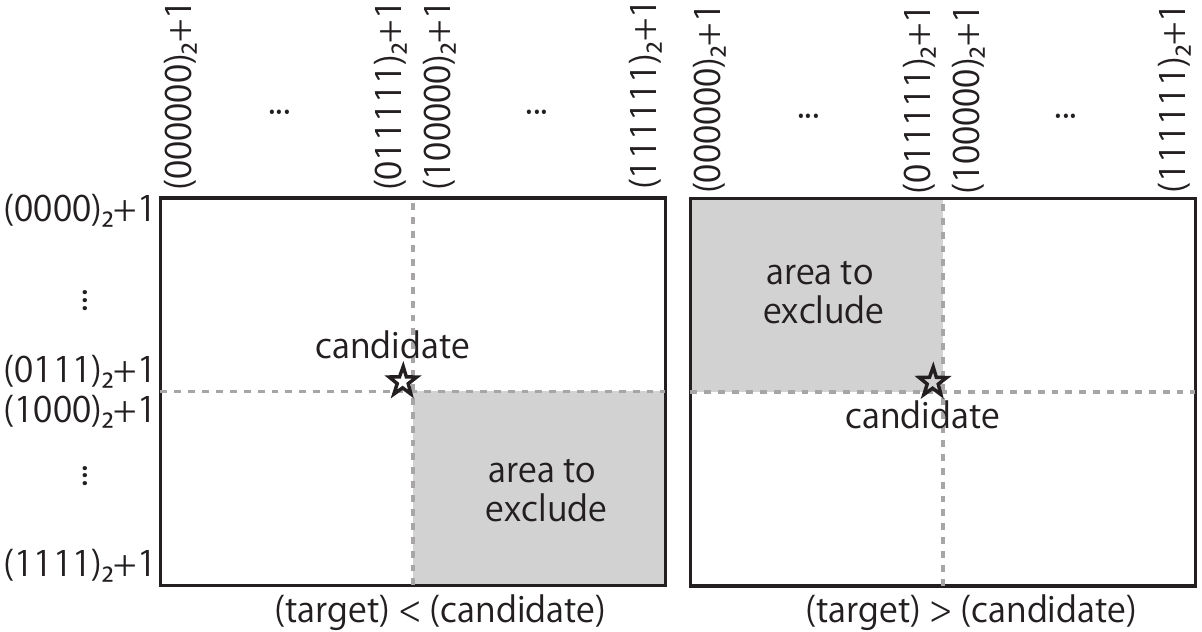}
\caption{Matrix $M_{\psi}$ in case when $\TDBS$ is first called.}
\label{fig:assignment_area_for_psi1} 
\end{center}
\end{figure}

In this subsubsection,
we fix $\psi_1$ to be $\psi_1^{(2)}$,
and for every $i \in [1, 10]$,
fix $z_i$ to be $z_i^{(2)}$.
After constructing $\psi_1$,
the algorithm constructs 
the set $\{(\psi_1, z_i)_4 \colon i \in [1, 10] \}$.
Then,
it searches the integer $\sum_{i=1}^{8} 4^{i-1}$ 
among the matrix $M_{\psi_1}$.
$\sum_{i=1}^{8} 4^{i-1}$ corresponds to a satisfying assignment for $\psi_1$.
An assignment ${\boldsymbol \sigma}$ satisfies $\psi_1$
if and only 
${\boldsymbol \sigma} \cdot \widehat{\boldsymbol \psi}_1$ is $(11111111)_4$;
i.e., $\sum_{i=1}^8 4^{i-1}$.
Needless to say,
representing all components in $M_{\psi_1}$
requires an exponential space for the input size.
However,
we can simultaneously do binary searches
in column and row directions in $M_{\psi_1}$
without explicitly constructing all the integers.
Let us describe more details of that search below.
Figure~\ref{fig:assignment_area_for_psi1} illustrates
the matrix $M_{\psi_1}$ in the first phase of the search.
We first set the assignment $011111111$ to the first candidate.
In Figure~\ref{fig:assignment_area_for_psi1},
the left and right squares represents $M_{\psi_1}$ in case when 
the value of the component corresponding to the assignment $11111111$
are smaller and larger than the one corresponding to the first candidate $01111111$.
respectively.

\newpage

\subsection{Formal details}
\label{subsec:details_of_algorithm}

In this subsection,
we describe the details of our algorithms for $\POS\OT\SAT$.

\begin{algorithm}[htbp] 
\SetAlgoNoLine%
\KwIn{A $3$CNF formula $\psi$}
\KwOut{A Boolean value,
which means whether $\psi$ is satisfiable.}
\Fn{$\ISSAT (\psi)$}{
$k_1 \leftarrow $ (the number of variables in $\psi$)
\;%
$m_1 \leftarrow $ (the number of clauses in $\psi$)
\;%
Rename variables in $\psi$ such that
$(\psi, z_1)_4 \leq \cdots \leq (\psi, z_{k_1})_4$
\;
\ForEach{$i \in [1, m_1]$}{
Sort literals in $C_i$
so that $C_i = z_{\ANBR{i, 3}} \lor z_{\ANBR{i, 2}} \lor z_{\ANBR{i, 1}}$
and $\ANBR{i, 3} > \ANBR{i, 2} > \ANBR{i, 1}$.}
\ForEach{$i \in [1, m_1]$}{
$C_{4i} \leftarrow (z_{k_1 + 3i} \lor z_{k_1 + 3i - 1} \lor z_{\ANBR{i,3}})$%
\tcp*[f]{a new clause}
\;
$C_{4i-1} \leftarrow (z_{k_1 + 3i} \lor z_{k_1 + 3i - 2} \lor z_{\ANBR{i,2}})$%
\tcp*[f]{a new clause}
\;
$C_{4i-2} \leftarrow (z_{k_1 + 3i} \lor z_{\ANBR{i,3}} \lor z_{\ANBR{i,2}})$%
\tcp*[f]{a new clause}
\;
$C_{4i-3} \leftarrow C_i$%
\tcp*[f]{a clause in $\psi$}
\;
}
$m \leftarrow 4 m_1$ 
\tcp*[f]{the number of clauses in $\varphi$}
\;
$\varphi \leftarrow C_m \land \cdots \land C_1$%
\;
$k_2 \leftarrow 3 m_1$ 
\tcp*[f]{the number of new variables}
\;%
Compute $\widehat{\boldsymbol \varphi}$.
\;
$t \leftarrow \sum_{i=1}^{m} 4^{i-1}$.
\tcp*[f]{a target integer}
\;%
$\mathbf{p} \leftarrow (0, 0)$
\tcp*[f]{the smallest corner in the search area}
\;%
$\mathbf{q} \leftarrow (2^{k_1}-1, 2^{k_2}-1)$
\tcp*[f]{the largest corner in the search area}
\;%
\KwRet{$\TDBS(\widehat{\boldsymbol \varphi}, \mathbf{p}, \mathbf{q}, k_1, t)$}
\;%
} 
\caption{$\ISSAT$}\label{alg:1}
\end{algorithm}

\begin{algorithm}[htbp] 
\SetAlgoNoLine%
\KwIn{A $5$-tuple 
$(\widehat{\boldsymbol \varphi}, \mathbf{p}, \mathbf{q}, k_1, t)$,
where $\widehat{\boldsymbol \varphi}$ is in $\mathbb{N}^+$;
$\mathbf{p} \in \mathbb{N}_+^2$ and $\mathbf{q} \in \mathbb{N}_+^2$ 
mean the smallest and largest points in the area to search;
$k_1$ means a bias for distinguishing coordinates;
and $t \in \mathbb{N}_+$ means a target value.%
}
\KwOut{A Boolean value,
which means whether
$t$ can be represented as $\mathbf{b} \cdot \widehat{\boldsymbol \varphi}$ 
for some $\mathbf{b} \in \{0, 1\}^k$.%
}

\Fn{$\TDBS(\widehat{\boldsymbol \varphi}, \mathbf{p}, \mathbf{q}, k_1, t)$}{
\If{$q_1 - p_1 \leq 1$ and $q_2 - p_2 \leq 1$}{
\KwRet{$0$}
}
$k \leftarrow $ (the number of components of $\widehat{\boldsymbol \varphi}$)
\;
$r \leftarrow 2^{k_1} \FLOOR{\frac{p_1 + q_1}{2}} + \FLOOR{\frac{p_2 + q_2}{2}}$%
\tcp*[f]{$l$ means an indicator for $\widehat{\boldsymbol \varphi}$.}
\;
$s \leftarrow \sum_{j=1}^{k} (\varphi, z_j)_4 \left(\FLOOR{r / 2^{j-1}} \MOD 2\right)$%
\tcp*[f]{$s$ means a candidate.}
\;
\If(\tcp*[f]{(the target) = (the current candidate)}){$s = t$}{%
\KwRet{$1$}%
}%
\ElseIf(\tcp*[f]{(the target) < (the current candidate)}){$t < s$}{%
\If{$q_1 - p_1 \geq 2$
and $q_2 - p_2 \geq 2$}{%
\If{$\TDBS(\widehat{\boldsymbol \varphi}, 
(p_1, \FLOOR{\frac{p_1 + q_1}{2}}),
(p_2, \FLOOR{\frac{p_2 + q_2}{2}}), 
k_1,
t) = 1$}{%
\KwRet{$1$}
}
\If{$\TDBS(\widehat{\boldsymbol \varphi}, 
(p_1, \FLOOR{\frac{p_1 + q_1}{2}}),
(\CEIL{\frac{p_2 + q_2}{2}}, q_2), 
k_1,
t)= 1$}{%
\KwRet{$1$}
}
\KwRet{$\TDBS(\widehat{\boldsymbol \varphi}, 
(\CEIL{\frac{p_1 + q_1}{2}}, q_1),
(p_2, \FLOOR{\frac{p_2 + q_2}{2}}), 
k_1,
t)$}
}
\ElseIf{$q_1 - p_1 \leq 1$}{%
\KwRet{$\TDBS(\widehat{\boldsymbol \varphi}, 
(p_1, q_1),
(p_2, \FLOOR{\frac{p_2 + q_2}{2}}), 
k_1,
t)$}
}
\Else(\tcp*[f]{when $q_2 - p_2 \leq 1$}){%
\KwRet{$\TDBS(\widehat{\boldsymbol \varphi}, 
(p_1, \FLOOR{\frac{p_2 + q_2}{2}}), 
(p_2, q_2),
k_1,
t)$}
}
}
\ElseIf(\tcp*[f]{(the target) > (the current candidate)}){$t > s$}{
\If{$q_1 - p_1 \geq 2$
and $q_2 - p_2 \geq 2$}{%
\If{$\TDBS(\widehat{\boldsymbol \varphi}, 
(p_1, \FLOOR{\frac{p_1 + q_1}{2}}),
(\CEIL{\frac{p_2 + q_2}{2}}, q_2), 
k_1,
t) = 1$}{%
\KwRet{$1$}
}
\If{$\TDBS(\widehat{\boldsymbol \varphi}, 
(\CEIL{\frac{p_1 + q_1}{2}}, q_1),
(p_2, \FLOOR{\frac{p_2 + q_2}{2}}), 
k_1,
t)= 1$}{%
\KwRet{$1$}
}
\KwRet{$\TDBS(\widehat{\boldsymbol \varphi}, 
(\CEIL{\frac{p_1 + q_1}{2}}, q_1),
(\CEIL{\frac{p_2 + q_2}{2}}, q_2), 
k_1,
t)$}
}
\ElseIf{$q_1 - p_1 \leq 1$}{%
\KwRet{$\TDBS(\widehat{\boldsymbol \varphi}, 
(p_1, q_1),
(\CEIL{\frac{p_2 + q_2}{2}}, q_2), 
k_1,
t)$}
}
\Else(\tcp*[f]{when $q_2 - p_2 \leq 1$}){%
\KwRet{$\TDBS(\widehat{\boldsymbol \varphi}, 
(\CEIL{\frac{p_1 + q_1}{2}}, q_1), 
(p_2, q_2),
k_1,
t)$}
}
}
}
\caption{$\TDBS$}\label{alg:3}
\end{algorithm}

\newpage

\subsection{Validity}
\label{subsec:validity_of_algorithm}

We fix $\widehat{\boldsymbol \varphi}$ to be as in Algorithm~\ref{alg:1}.
We fix $f$ to be a mapping such that
$f({\boldsymbol \sigma})$ is 
the pair $(({\boldsymbol \sigma}_{1:k_1})^R_2+1, ({\boldsymbol \sigma}_{k_1+1:k})^R_2+1)$
for a given assignment ${\boldsymbol \sigma}$.
We fix $M$ to be the $(2^{k_1}, 2^{k_2})$-matrix
whose $(i, j)$-element is $f^{-1}(i, j) \cdot \widehat{\boldsymbol \varphi}$,
where $i \in [1, 2^{k_1}]$ and $j \in [1, 2^{k_2}]$.
Lemma~\ref{lem:1} is necessary
for $\TDBS$ to execute its procedure expectedly.
Note that 
we do not compute all parts of the matrix $M$
in Algorithms~\ref{alg:1}.

\begin{lemma}[$2$-dimensional sortability]
\label{lem:1}
Let $x$ and $y$ be integers 
in $[2, 2^{k_1}]$ and $[2, 2^{k_2}]$, 
respectively.
Then, the following holds.
\begin{enumerate}[(\ref{lem:1}.1)]
\item\label{enum:l1-1}
$m_{x-1, y} < m_{x, y}$.
\item\label{enum:l1-2}
$m_{x, y-1} < m_{x, y}$.
\end{enumerate}
\end{lemma}
\PROOFSTARTX{of Lemma~\ref{lem:1}}
We fix ${\boldsymbol \xi}$ to be the vector $f^{-1}(x, y)$.

Let us first prove the inequality (\ref{lem:1}.\ref{enum:l1-1}).
Let ${\boldsymbol \theta}$ be the vector $f^{-1}(x-1, y)$.
By definition,
$({\boldsymbol \xi}^{\mathrm{R}})_2 = 2^{k_1} y + x$
and $({\boldsymbol \theta}^{\mathrm{R}})_2 = 2^{k_1} y + x - 1$.
Then, there is an integer $l_v \in [1, k_1]$
such that
$\theta_i = \xi_i$ for every $i \in [l_v+1, k_1+k_2]$;
$\theta_{l_v} = 0$; and $\xi_{l_v} = 1$.
Then, the following claim 
implies that
${\boldsymbol \xi} \cdot \widehat{\boldsymbol \varphi} > {\boldsymbol \theta} \cdot \widehat{\boldsymbol \varphi}$;
i.e., $m_{x, y} > m_{x-1, y}$.
\begin{claim}
\label{cla:4} 
$(\varphi, z_{l_v})_4 > \sum_{i=1}^{l_v-1} (\varphi, z_i)_4$.
\end{claim}
\PROOFSTARTX{of Claim~\ref{cla:4}}
The proof is by induction on $l_v$.
Let $l_c = \max\{j \colon z_{l_v} \in C_j, j \in [1, m_1]\}$.
Let $l_b$ be an integer in $[1, 3]$
such that $z_{\ANBR{l_c, l_b}} \in C_{l_c}$.
First, suppose that $l_b = 1$.
By lines $4$-$6$ in Algorithm~\ref{alg:1},
$(\varphi, z_{\ANBR{l_c, l_b} - 1}) \leq (\varphi, z_{\ANBR{l_c-1, 3}})$.
By lines $8$-$11$ in Algorithm~\ref{alg:1},
$2(\varphi, z_{\ANBR{l_c-1, 3}})_4 < (\varphi, z_{\ANBR{l_c, l_b}})_4$.
Thus,
$2(\varphi, z_{\ANBR{l_c, l_b}-1})_4 < (\varphi, z_{\ANBR{l_c, l_b}})_4$.
Next, suppose that $l_b \in \{2, 3\}$.
By lines $4$-$6$ in Algorithm~\ref{alg:1},
$(\varphi, z_{\ANBR{l_c, l_b} - 1}) = (\varphi, z_{\ANBR{l_c, l_b-1}})$.
By lines $8$-$11$ in Algorithm~\ref{alg:1},
$2(\varphi, z_{\ANBR{l_c, l_b-1}})_4 < (\varphi, z_{\ANBR{l_c, l_b}})_4$.
Thus,
$2(\varphi, z_{\ANBR{l_c, l_b}-1})_4 < (\varphi, z_{\ANBR{l_c, l_b}})_4$.
By induction hypothesis,
$\sum_{i=1}^{\mathrm{max}_v-2} (\varphi, z_i)_4 < (\varphi, z_{\mathrm{max}_v-1})_4$.
Thus,
$\sum_{i=1}^{\mathrm{max}_v-1} (\varphi, z_i)_4 < 2(\varphi, z_{\mathrm{max}_v-1})_4$.
Consequently,
$\sum_{i=1}^{\mathrm{max}_v-1} (\varphi, z_i)_4 < (\varphi, z_{\mathrm{max}_v})_4$.
\QED (Claim)

Let us next prove the inequality~(\ref{lem:1}.\ref{enum:l1-2}).
Let ${\boldsymbol \eta}$ be the vector $f^{-1}(x, y-1)$.
By definition,
$({\boldsymbol \xi}^{\mathrm{R}})_2 = 2^{k_1} y + x$
and $({\boldsymbol \eta}^{\mathrm{R}})_2 = 2^{k_1} (y-1) + x$.
By line~$14$ in Algorithm~\ref{alg:1},
$k_2 = 3 m_1$.
There are integers $l_0 \in [1, m_1]$ and $l_1 \in [0, 2]$
such that
$\eta_i = \xi_i$ for every $i \in [k_1+3l_0-l_1-1, k_1+k_2]$;
$\theta_{k_1 + 3l_0-l_1} = 0$; 
and $\xi_{k_1 + 3 l_0-l_1} = 1$.
Let $\lambda =  k_1 + 3 l_0 - l_1$. 
Then, the following claim implies that
${\boldsymbol \xi} \cdot \widehat{\boldsymbol \varphi} > {\boldsymbol \eta} \cdot \widehat{\boldsymbol \varphi}$;
i.e., $\lambda_{x, y} \geq \lambda_{x-1, y}$.

\begin{claim}
\label{cla:5} 
$(\varphi, z_\lambda)_4 > \sum_{i=k_1+1}^{\lambda - 1} (\varphi, z_i)_4$.
\end{claim}
\PROOFSTARTX{of Claim~\ref{cla:5}}
The proof is by induction on $\lambda$.
In Algorithm~\ref{alg:1},
operations for $z_\lambda$ 
are only in the loop of lines $7$-$11$.
Moreover, 
those operations are only in time 
when $i = l_0$
during all iterations of that loop.
Thus, by lines $8$-$11$ in Algorithm~\ref{alg:1},
$(\varphi, z_{k_1 + 3 l_0 - l_1})_4
> \sum_{i=l_1+1}^2 (\varphi, z_{k_1 + 3 l_0 - i})_4 
+ 2 (\varphi, z_{k_1 + 3 (l_0-1)})_4$. 
That is,
$(\varphi, z_\lambda)_4
> \sum_{i=k_1 + 3 l_0 - 2}^{\lambda-1} (\varphi, z_i)_4 
+ 2 (\varphi, z_{k_1 + 3 (l_0-1)})_4$. 
By induction hypothesis,
$(\varphi, z_{k_1 + 3 (l_0-1)})_4 > \sum_{i=1}^{k_1 + 3 (l_0-1) - 1} (\varphi, z_i)_4$.
Thus,
$2 (\varphi, z_{k_1 + 3 (l_0-1)})_4 > \sum_{i=1}^{k_1 + 3 (l_0-1)} (\varphi, z_i)_4$.
Consequently,
$(\varphi, z_\lambda)_4 > \sum_{i=1}^{\lambda-1} (\varphi, z_i)_4$.%
\QED (Claim~\ref{cla:5})%
\QED (Lemma~\ref{lem:1})

\begin{lemma}[Equivalence of formulae]
\label{lem:2}
Let $\psi$ and $\varphi$ be as in Algorithm~\ref{alg:1}.
Then, $\psi$ is satisfiable if and only if
$\varphi$ is satisfiable.
\end{lemma}
\PROOFSTARTX{of Lemma~\ref{lem:2}}
By line~$11$ in Algorithm~\ref{alg:1},
$\varphi$ contains all clauses in $\psi$. 
Thus, the {\itshape ``if''} part is trivial.
We will prove the {\itshape ``only if''} part below.
Suppose that $\psi$ is satisfiable.
Let ${\boldsymbol \sigma}$ be a satisfying assignment for $\psi$.
It suffices to show that 
there is a satisfying assignment ${\boldsymbol \mu}$ for $\varphi$ such that
$\mu_i = \sigma_i$ for every $i \in [1, k_1]$.
Let $j \in [1, m_1]$.
By line~$11$ in Algorithm~\ref{alg:1},
$C_{4j-3}$ in $\varphi$ is equal to $C_j$ in $\psi$.
Thus, ${\boldsymbol \mu}$ satisfies $C_{4j-3}$ in $\varphi$. 
Let us then consider the conjunction of the clauses 
$C_{4j} = (z_{k_1 + 3j} \lor z_{k_1 + 3j - 1} \lor z_{\ANBR{j,3}})$,
$C_{4j-1} = (z_{k_1 + 3j} \lor z_{k_1 + 3j - 2} \lor z_{\ANBR{j,2}})$, 
and $C_{4j-2} = (z_{k_1 + 3j} \lor z_{\ANBR{j,3}} \lor z_{\ANBR{j,2}})$ 
in $\varphi$.
In Algorithm~\ref{alg:1},
for every $[0, 2]$,
$z_{k_1 + 3j - l}$ occurs only in 
$C_{4j} \land C_{4j-1} \land C_{4j-2}$ in $\varphi$.
Thus, we can assign 
$z_{k_1 + 3j}$, $z_{k_1 + 3j - 1}$, and $z_{k_1 + 3j - 2}$
to values without affecting
the values of all clauses except for $C_{4j}$, $C_{4j-1}$, and $C_{4j-2}$.
First, suppose that 
$z_{\ANBR{j,1}}$, $z_{\ANBR{j,2}}$, and $z_{\ANBR{j,3}}$ 
are assigned $1$, $0$, and $0$ in $\psi$, respectively.
If ${\boldsymbol \mu}$ assigns 
$z_{k_1 + 3j}$, $z_{k_1 + 3j - 1}$, and $z_{k_1 + 3j - 2}$ 
to $1$, $0$, and $0$, respectively,
then ${\boldsymbol \mu}$ satisfies $C_{4j}$, $C_{4j-1}$, and $C_{4j-2}$. 
Next, suppose that 
$z_{\ANBR{j,1}}$, $z_{\ANBR{j,2}}$, and $z_{\ANBR{j,3}}$ 
are assigned $0$, $1$, and $0$ in $\psi$, respectively.
If ${\boldsymbol \mu}$ assigns 
$z_{k_1 + 3j}$, $z_{ k_1 + 3j - 1}$, and $z_{ k_1 + 3j - 2}$ 
to $0$, $1$, and $0$, respectively,
then ${\boldsymbol \mu}$ satisfies $C_{4j}$, $C_{4j-1}$, and $C_{4j-2}$. 
Next, suppose that
$z_{\ANBR{j,1}}$, $z_{\ANBR{j,2}}$, and $z_{\ANBR{j,3}}$ 
are assigned $0$, $0$, and $1$ in $\psi$, respectively.
If ${\boldsymbol \mu}$ assigns 
$z_{k_1 + 3j }$, $z_{k_1 + 3j- 1}$ and $z_{k_1 + 3j - 2}$ 
to $0$, $0$, and $1$, respectively,
then ${\boldsymbol \mu}$ satisfies $C_{4j}$, $C_{4j-1}$, and $C_{4j-2}$. 
Consequently,
$\varphi$ is satisfiable.
\QED

\subsection{Running time}
\label{subsec:running_time_of_algorithm}

In this subsection, 
we analyze the running time of the proposed algorithm.

\begin{lemma}[Polynomial-time computability]
\label{lem:3}
Given $\psi$, $\ISSAT(\psi)$ runs in time polynomial in $k_1$.
\end{lemma}
\PROOFSTARTX{of Lemma~\ref{lem:3}}%
Each operation of addition, subtraction, multiplication, division, 
mod, floor, and ceiling
can be computed in time polynomial in $k_1$.
Those operations are executed $H(k,m)$ times,
where $H$ is a linear function of $k$ and $m$.
In lines $2$-$18$ in Algorithm~\ref{alg:1},
we spend time as follows.
Note that $m_1 = \Theta(k_1)$.

In line~$1$, 
we can count the variables in $\psi$ 
in time linear in $k_1$.
In line~$2$ in Algorithm~\ref{alg:1},
we count the clause in $\psi$ in time linear in $m_1$.
For every $j \in [1, k_1]$,
we can compute $(\psi, z_j)_4$ in time polynomial in $k_1$.
Moreover,
we can sort the sequence
$(\psi, z_1)_4, \cdots, (\psi, z_{k_1})_4$
in ascending order in time polynomial in $k_1$.
After that procedure, 
we can rename variables in $\psi$
$(\psi, z_1)_4 < \cdots < (\psi, z_{k_1})_4$
in time linear in $k_1$.
That is, we can do the step in line~$4$
in time polynomial in $k_1$.
For every $j \in [1, m_1]$,
we can do the step in line~$6$ in time linear in $k_1$.
Thus,
we can do the loop in lines~$5$-$6$ in time polynomial in $k_1$.
For every $j \in [1, m_1]$,
we can do the step in lines~$8$-$11$ in time linear in $k_1$.
Thus,
we can do the loop in lines~$7$-$11$ in time polynomial in $k_1$.
In line~$15$,
we compute $(\varphi, z_j)_4$ for every $j \in [1, k_1+3m_1]$.
That procedure are done in time polynomial in $k_1$.
By line $12$,
$m = 4m_1$.
Thus, $m = \Theta(k_1)$.
In line~$16$,
we can compute $\sum_{i=1}^m 4^{i-1}$ in time polynomial in $k_1$.
By line $14$,
$k_2 = 3m_1$.
Thus, $k_2 = \Theta(k_1)$.
In line~$18$,
we can compute $2^{k_1}$ and $2^{k_2}$ in time polynomial in $k_1$.
Moreover, by Claim~\ref{cla:1} below,
we can do the step in line~$19$ in time polynomial in $k_1$.
Consequently,
the total running time of $\ISSAT(\psi)$ is $O(k_1)$.

\begin{claim}
\label{cla:1}
$\TDBS(\widehat{\boldsymbol \varphi}, \mathbf{p}, \mathbf{q}, k_1, t)$ 
runs in time polynomial in $k_1$.
\end{claim}
\PROOFSTARTX{of Claim~\ref{cla:1}} 
Let us first analyze the running time of $\TDBS$ 
for all steps except for recursive calls.
By the above discussion,
the number $k$ of components in $\widehat{\boldsymbol \varphi}$
is $O(k_1)$.
In line~$4$,
we can count the number of components in $\widehat{\boldsymbol \varphi}$
in time polynomial in $k_1$.
We can compute the expression in the righthand side
in line~$5$.
$r$ takes at most $O(k_1)$ bits.
The variable $s$ has the largest bit length
in all variables in Algorithm~\ref{alg:3}.
Its bit length is at most $O(k_1)$.
Thus,
all steps in lines~2-30
except for recursive calls
can be executed in time polynomial in $k_1$.

In every recursive call,
the bit lengths of the first, fourth, and fifth arguments
are the same as the one in the calling procedure;
and moreover, the ones of the second and third arguments
are about the halves of the one in the calling procedure.
We denote the sum of the bit lengths for representing 
$\widehat{\boldsymbol \varphi}$, $k_1$, and $t$
by $\lambda_0$.
We denote the sum of the bit lengths for representing 
$\mathbf{p}$ and $\mathbf{q}$
by $\lambda$.
In the first call for $\TDBS$,
$\lambda$ is $2k_1 + 2k_2$; i.e., $2k$.
The depth of recursion depends on $\lambda$, but independent of $\lambda_0$.
We define $T_0(\lambda_0)$ as an upper bound for the time 
of all steps except for recursive calls
in $\TDBS(\widehat{\boldsymbol \varphi}, \mathbf{p}, \mathbf{q}, k_1, t)$.
We define $T(\lambda_0, \lambda)$ as an upper bound for  
the total running time of
$\TDBS(\widehat{\boldsymbol \varphi}, \mathbf{p}, \mathbf{q}, k_1, t)$.

In a call of $\TDBS$,
there are the following six cases.
Let $s_{(1)}$ be the value of $s$ after line~6.
(\ref{lem:2}-I) $q_1 - p_1 \leq 1$ and 
$q_2 - p_2 \leq 1$;
i.e.,
the condition in line~2 is true.
(\ref{lem:2}-II) 
$t = s_{(1)}$,
$q_1 - p_1 \geq 2$,
and $q_2 - p_2 \geq 2$;
i.e.,
the condition in line~7 is true.
(\ref{lem:2}-III) 
$t < s_{(1)}$,
$q_1 - p_1 \geq 2$,
and $q_2 - p_2 \geq 2$;
i.e.,
the conditions in lines~9 and 10 are true.
(\ref{lem:2}-IV) 
$t < s_{(1)}$
and ($q_1 - p_1 \leq 1$ or $q_2 - p_2 \leq 1$);
i.e.,
the condition in lines~9 and 10 are true and false, respectively.
(\ref{lem:2}-V) 
$t > s_{(1)}$,
$q_1 - p_1 \geq 2$, 
and $q_2 - p_2 \geq 2$;
i.e.,
the conditions in lines~20 and 21 are true.
(\ref{lem:2}-VI) 
$t > s_{(1)}$
and ($q_1 - p_1 \leq 1$ or $q_2 - p_2 \leq 1$);
i.e.,
the condition in lines~20 and 21 are true and false, respectively.
Then, we obtain the following recurrence.
\begin{eqnarray}
T\left(\lambda_0, \lambda\right) = 
\begin{cases}
B(\lambda_0) & \text{in cases (\ref{lem:2}-I) and (\ref{lem:2}-II)}
\\
3 T\left(\lambda_0, \frac{\lambda}{4}\right) + T_0(\lambda_0) & \text{ in cases (\ref{lem:2}-III) or (\ref{lem:2}-V)}
\\
T\left(\lambda_0, \frac{\lambda}{2}\right) + T_0(\lambda_0) & \text{ in cases (\ref{lem:2}-IV) or (\ref{lem:2}-VI)},
\end{cases}
\label{eq:1}
\end{eqnarray}
where $B$ is a polynomial.
Let $T_1(\lambda_0) = B(\lambda_0) + T_0(\lambda_0)$.
Then, by Claim~\ref{cla:2} below,
$T(\lambda_0, \lambda)$ is of polynomial order in $\lambda$ and $\lambda_0$.
By the above discussion for Algorithm~\ref{alg:1},
$\lambda_0$ and $\lambda$ are of polynomial order in $k_1$.
Consequently,
$\TDBS(\widehat{\boldsymbol \varphi}, \mathbf{p}, \mathbf{q}, k_1, t)$
runs in time polynomial in $k_1$.
\QED (Claim~\ref{cla:1})

\begin{claim}
\label{cla:2}
$T(\lambda_0, \lambda) \leq \lambda^u T_1(\lambda_0)$, where $u = \frac{1}{\log_3 2}$.
\end{claim}
\PROOFSTARTX{of Claim~\ref{cla:2}}
The proof is by induction on $\lambda$.
By the definition of $T_1$,
$B(\lambda_0) < T_1(\lambda_0)$.
Thus,
in cases (\ref{lem:2}-I) and (\ref{lem:2}-II),
$T(\lambda_0, \lambda) \leq \lambda^u T_1(\lambda_0)$.
In cases (\ref{lem:2}-III) to (\ref{lem:2}-VI),
$T(\lambda_0, \lambda) \leq 3 T(\lambda_0, \lambda/4) + T_0(\lambda_0)$.
By induction hypothesis,
$T(\lambda_0, \lambda) \leq 3 (\lambda/4)^u T_1(\lambda_0) + T_0(\lambda_0)$.
By a rearrangement, 
$3 (\lambda/4)^u T_1(\lambda_0) + T_0(\lambda_0) 
= \lambda^u ((3/4^u)T_1(\lambda_0) + (1/\lambda^u)T_0(\lambda_0))$.
By the definition of $T_1$,
$T_0(\lambda_0) < T_1(\lambda_0)$.
By definition,
$\lambda \geq 4$.
Thus,
$(3/4^\lambda) + (1/\lambda^u) \leq 1$.
It follows that
$3 (\lambda/4)^u T_1(\lambda_0) + T_0(\lambda_0) \leq \lambda^u T_1(\lambda_0)$.
That is,
$T(\lambda_0, \lambda) \leq \lambda^u T_1(\lambda_0)$.
\QED (Claim~\ref{cla:2})%
\QED (Lemma~\ref{lem:3})


\begin{thebibliography}{1}

\bibitem{arora2009computational}
Sanjeev Arora and Boaz Barak.
\newblock {\em Computational Complexity: a Modern Approach}.
\newblock Cambridge University Press, New York, NY, 2009.

\bibitem{creignou2001complexity}
Nadia Creignou, Sanjeev Khanna, and Madhu Sudan.
\newblock {\em Complexity Classifications of Boolean Constraint Satisfaction
  Problems}.
\newblock SIAM, 2001.

\bibitem{Schaefer:1978:CSP:800133.804350}
Thomas~J. Schaefer.
\newblock The complexity of satisfiability problems.
\newblock In {\em Proceedings of the Tenth Annual ACM Symposium on Theory of
  Computing}, STOC '78, pages 216--226, New York, NY, USA, 1978. ACM.

\bibitem{Wegener:1987:CBF:35517}
Ingo Wegener.
\newblock {\em The Complexity of Boolean Functions}.
\newblock John Wiley \& Sons, Inc., New York, NY, USA, 1987.

\end{thebibliography}
\end{document}